\documentclass[aps,prb,showpacs,superscriptaddress]{revtex4}
\usepackage{latexsym}
\usepackage{amssymb}
\usepackage[dvips]{graphicx}
\begin{document}
\title{Long range scattering effects on spin Hall current in $p$-type bulk semiconductors}
\author{S. Y. Liu}
\email{liusy@mail.sjtu.edu.cn}
\affiliation{Department of Physics and Engineering Physics, Stevens
Institute of Technology, Hoboken, New Jersey 07030, USA}
\affiliation{Department of Physics, Shanghai Jiaotong University, 1954
Huashan Road, Shanghai 200030, China}
\author{ Norman J. M. Horing}
\affiliation{Department of Physics and Engineering Physics, Stevens
Institute of Technology, Hoboken, New Jersey 07030, USA}
\author{X. L. Lei}
\affiliation{Department of Physics, Shanghai Jiaotong University, 1954
Huashan Road, Shanghai 200030, China}

\date{\today}
\begin{abstract}
Employing a nonequilibrium Green's function approach, we examine
the effects of long-range hole-impurity scattering on
spin-Hall current in $p$-type bulk semiconductors within the
framework of the self-consistent Born approximation. We find that,
contrary to the null effect of short-range scattering on spin-Hall
current, long-range collisions do produce a nonvanishing
contribution to the spin-Hall current, which is independent of
impurity density in the diffusive regime and relates only to hole
states near the Fermi surface. The sign of this contribution is
opposite to that of the previously predicted disorder-independent
spin-Hall current, leading to a sign change of the total spin-Hall
current as hole density varies. Furthermore, we also make clear
that the disorder-independent spin-Hall effect is a result of an
interband polarization directly induced by the dc electric field
with contributions from all hole states in the Fermi sea.
\end{abstract}

\pacs{72.10.-d, 72.25.Dc, 72.25.-b}

 \maketitle
\section{Introduction}

Recently, there have been extensive studies of the physics of the
spin-orbit (SO) interaction in condensed matter. The most
intriguing phenomenon induced by SO coupling is the spin-Hall
effect (SHE): when a dc electric field is applied, the SO
interaction may result in a net nonvanishing spin current flow
along the transverse direction. The SHE is classified into two
types according to its origin, an {\it extrinsic} spin-orbit
Hamiltonian term induced by carrier-impurity scattering
potentials\cite{DP,HS} and an {\it intrinsic} spin-orbit
Hamiltonian term arising from free carrier
kinetics.\cite{Zhang,Sinova} The intrinsic spin-Hall effect was
originally thought to be independent of carrier-impurity
scattering. Experimentally, the SHE was observed in a $n$-type
bulk semiconductor\cite{Kato} and in a two-dimensional (2D)
heavy-hole system.\cite{Wunderlich}

However, further studies have indicated that the spin-Hall effect
associated with the intrinsic mechanism can be strongly affected
by carrier-impurity scattering
(disorder).\cite{Loss1,Nomura,Loss2,Dimitrova,Bauer,Raimondi,Khaetskii,Halperin,Liu1,Nagaosa,Liu2,Bernevig,
Liu3,Halperin2,Murakami,Chen} (To avoid confusion, we use the term
"intrinsic SHE" to refer to the
total spin-Hall effect arising from the SO coupling terms of the
Hamiltonian that do not explicitly involve scattering; ultimately,
this is corrected by scattering, but the part that is unaffected by scattering
will be termed the intrinsic "disorder-independent"
SHE.) In diffusive 2D semiconductors, there always exists a
contribution to the intrinsic spin-Hall current which arises from
spin-conserving electron-impurity scattering, but it is
independent of impurity density within the diffusive regime. For
2D {\it electron} systems with Rashba SO coupling, this
disorder-related spin-Hall current leads to the vanishing of the
total intrinsic spin-Hall current, irrespective of the specific
form of the scattering potential, of the collisional broadening,
and of temperature.\cite{Liu2} In 2D Rashba {\it heavy-hole}
systems, disorder affects the intrinsic SHE in a different
fashion: contributions from short-range collisions to the SHE
vanish,\cite{Bernevig} while long-range electron-impurity
scattering produces a nonvanishing disorder-related spin-Hall
current, whose sign changes with variation of the hole
density.\cite{Liu3,Halperin2}

To date, the effect of disorder on the intrinsic spin-Hall current
in $p$-type bulk semiconductors has been studied relatively little.
Employing a Kubo formula, Murakami found a null disorder effect on
the intrinsic SHE for short-range hole-impurity
collisions.\cite{Murakami} The crossover of the SHE from the
diffusive to the hopping regime has been investigated by modeling
finite-size samples (with a maximum of $50 \times 50\times 50$
lattice sites) by Chen, {\it et al}.\cite{Chen} In this paper, we employ a
nonequilibrium Green's function approach to study the effect of
more realistic {\it long-range} hole-impurity scattering on
the intrinsic spin-Hall current in a diffusive $p$-type bulk
semiconductor. We find that, in such a system, the contribution of
hole-impurity collisions to the intrinsic spin-Hall current is
finite and it is independent of impurity density within the
diffusive regime. Moreover, this disorder contribution has its
sign opposite to that of the disorder-independent one, leading to
a sign change of the total spin-Hall current as the hole density
varies. Furthermore, we make clear that the
disorder-independent spin-Hall effect arises from an interband
polarization process directly induced by the dc electric field and
it involves all hole states below the Fermi surface. In contrast
to this, the disorder contribution to the intrinsic SHE originates
from a disorder-mediated polarization between two hole bands and
is associated only with hole states in the vicinity of the Fermi
surface. Also, we numerically examine the hole-density
dependencies of the spin-Hall conductivity and mobility.

This paper is organized as follows. In Sec. II, we derive the
kinetic equation for the nonequilibrium distribution function and
discuss the origins of the disorder-independent and
disorder-related spin-Hall currents. In Sec. III, we perform a
numerical calculation to investigate the effect of long-range
hole-impurity scattering on the spin-Hall current. Finally, we
review our results in Sec. IV.

\section{Formalism}

\subsection{Kinetic equation}
It is well known that for semiconductors with diamond structure
(e.g. Si, Ge) or zinc blende structure (e.g. GaAs), the tops of
the valence bands usually are split into fourfold degenerate
$S=3/2$ and twofold degenerate $S=1/2$ states due to the
spin-orbit interaction ($S$ denotes the total angular momentum of
the atomic orbital). Near the top of the $S=3/2$ valence bands,
the electronic structure can be described by a simplified Luttinger
Hamiltonian\cite{Luttinger}
\begin{equation}
{\check h}_0({\bf p})=\frac 1{2m}\left [\left (\gamma_1+\frac 52 \gamma_2 \right )p^2-
2\gamma_2({\bf p}\cdot {\bf S})^2\right],\label{ham}
\end{equation}
where, ${\bf p}\equiv (p_x,p_y,p_z)\equiv (p\sin\theta_{\bf
p}\cos\phi_{\bf p} ,p\sin\theta_{\bf p}\sin\phi_{\bf
p},p\cos\theta_{\bf p})$ is the three-dimensional (3D) hole
momentum, $m$ is the free electron mass, ${\bf S}\equiv
(S_x,S_y,S_z)$ are the spin-$3/2$ matrices, and $\gamma_1$ and
$\gamma_2$ are the material constants. (As in previous
studies,\cite{Zhang,Murakami,Chen,Zhang2} we simplified by setting
$\gamma_3=\gamma_2$ in the original Luttinger Hamiltonian
presented in Ref.\,\onlinecite{Luttinger}).

By a local unitary spinor transformation, $ U_{\bf
p}=\exp({-iS_z\phi_{\bf p}})\exp({-iS_y\theta_{\bf p}})$,
Hamiltonian (\ref{ham}) can be diagonalized as ${\hat h}_0({\bf
p}) = U^+_{\bf p}\check h_0({\bf p}) U_{\bf p}={\rm
diag}[\varepsilon_{H}(p), \varepsilon_{L}(p), \varepsilon_{L}(p),
\varepsilon_{H}(p)]$. Here, $\varepsilon_{H}
(p)=\frac{\gamma_1-2\gamma_2}{2m}p^2$ and $\varepsilon_{L}
(p)=\frac{\gamma_1+2\gamma_2}{2m}p^2$ are, respectively, the
dispersion relations of the heavy- and light-hole bands.
Physically, this transformation corresponds to a change from a
spin basis to a helicity basis.

In a realistic 3D system, holes experience scattering by
impurities. We assume that this interaction between holes and
impurities can be characterized by an isotropic potential,
$V(|{\bf p}-{\bf k}|)$, which corresponds to scattering a hole
from state ${\bf p}$ to state ${\bf k}$. In the helicity basis,
the scattering potential takes the transformed form, $\hat T ({\bf
p},{\bf k})=U^+_{\bf p}V(|{\bf p}-{\bf k}|)U_{\bf k}$.

We are interested in the spin-Hall current in a bulk hole system
driven by a dc electric field ${\bf E}$ along the $z$ axis. In
Coulomb gauge, this electric field can be described by the scalar
potential, $V\equiv -e{\bf E}\cdot {\bf r}$, with ${\bf r}$ as the
hole coordinate. Without loss of generality, we specifically study
a spin current, $J_{y}^{x}$, that is polarized along the $x$ axis
and flows along the $y$ axis. In the spin basis, the conserved
single-particle spin-Hall operator, ${\check j}_{y}^{x}$, is
defined as\cite{Zhang2}
\begin{equation}
 \check j_{y}^{x}({\bf p})
 =\frac 16 \left \{\frac{\partial \check h_0}{\partial p_y}, P_{\bf p}^LS_xP_{\bf p}^L+P_{\bf p}^HS_xP_{\bf p}^H\right
 \},
 \end{equation}
with $P_{\bf p}^L$ and $P_{\bf p}^H$, respectively, as projection operators
onto the states of light- and heavy-hole bands: $P_{\bf p}^L=\frac
98 -\frac{1}{2p^2}({\bf p}\cdot {\bf S})^2$, $P_{\bf p}^H=1-P_{\bf
p}^L$. Taking a statistical ensemble average, the observed net
spin-Hall current is given by
\begin{equation}
J_{y}^{x}=\sum_{\bf p}{\rm Tr}[\check j_{y}^{x}({\bf p})\check\rho({\bf p})],
\end{equation}
where $\check \rho({\bf p})$ is the distribution function related
to the nonequilibrium "lesser" Green's function, ${\check {\rm
G}}^<({\bf p},\omega)$, as given by $\check \rho({\bf p})=-i\int \frac
{d\omega}{2\pi} \check {\rm G}^{<}({\bf p},\omega)$. Also, $J_y^x$
can be determined in helicity basis via
\begin{equation}
J_y^x=\sum_{{\bf p}}{\rm Tr}[{\hat j}_y^x({\bf p}){\hat \rho}({\bf p})],\label{JYX}
\end{equation}
with ${\hat j}_y^x({\bf p})=U_{{\bf p}}^+\check j_y^x({\bf
p})U_{{\bf p}}$ and ${\hat \rho}({\bf p}) =U^+({\bf p})\check
\rho({\bf p}) U({\bf p})$ being the helicity-basis
single-particle spin current operator and distribution function,
respectively. Explicitly, Eq.\,(\ref{JYX}) can be rewritten as
($\hat \rho_{\mu\nu}({\bf p})$ are the matrix elements of $\hat \rho({\bf p})$
in helicity basis; $\mu,\nu=1,2,3,4$)
\begin{eqnarray}
J_y^x&=&\frac{\sqrt{3}\gamma_2}{m}\sum_{\bf p}p\left \{4\cos^2\phi_{\bf p}\sin \theta_{\bf p} {\rm Im}[\hat \rho_{12}({\bf p})+\hat \rho_{34}({\bf p})]
\right.\nonumber\\
&&-\sin(2\phi_{\bf p})\sin(2 \theta_{\bf p}) {\rm Re}[\hat \rho_{12}({\bf p})+\hat \rho_{34}({\bf p})]
\nonumber\\
&&
+2\cos(2 \phi_{\bf p})\cos\theta_{\bf p}{\rm Im}[\hat \rho_{13}({\bf p})-\hat \rho_{24}({\bf p})]\nonumber\\
&&
\left .-\sin(2\phi_{\bf p})[1+\cos^2 \theta_{\bf p}] {\rm Re}[\hat \rho_{13}({\bf p})-\hat \rho_{24}({\bf p})]
\right \}.\label{JYXE}
\end{eqnarray}
Here, the Hermitian property of the distribution function, {\it
i.e.} $\hat {\rho}({\bf p})=\hat {\rho}^+({\bf p})$, has been
used. It is clear from Eq.\,(\ref{JYXE}) that contributions to the
spin-Hall current arise only from those elements of the
distribution function which describe the interband polarization,
such as $\hat \rho_{12}(\bf p)$, $\hat \rho_{13}(\bf p)$, $\hat
\rho_{34}(\bf p)$ and $\hat \rho_{24}(\bf p)$. The vanishing of
spin-Hall current contributions from the diagonal elements of the distribution
function is associated with the helicity degeneracy of the hole
bands in $p$-type bulk semiconductors. The diagonal elements of
the distribution function for holes in same band but with opposite
helicities are the same, {\it i.e.} $\hat \rho_{22}({\bf
p})=\rho_{33}({\bf p})$ and $\hat \rho_{11}({\bf
p})=\rho_{44}({\bf p})$. However, the corresponding diagonal
elements of the single-particle spin current have opposite signs
due to opposite helicities, $(\hat j_{y}^x)_{22}({\bf p})=-(\hat
j_y^x)_{33}({\bf p})$ and $(\hat j_{y}^x)_{11}({\bf p})=-(\hat
j_y^x)_{44}({\bf p})$. As a result, the net contributions to
spin-Hall current from the diagonal elements of distribution
function are eliminated.

In order to carry out the calculation of spin-Hall current, it is
necessary to determine the hole distribution function.\cite{Wu} Under
homogeneous and steady-state conditions, the spin-basis
distribution, ${\check \rho}({\bf p})$, obeys a kinetic equation
taken the form,
\begin{equation}
e{\bf E}\cdot [\nabla_{\bf p} {\check \rho}({\bf p})]+i[{\check
h}_0({\bf p}),{\check \rho}({\bf p})]=-\check I,\label{KE}
\end{equation}
with $\check I$ as a collision term given by
\begin{equation}
\check I= \int \frac{d\omega}{2\pi}({\check \Sigma}^r_{\bf
p}{\check {\rm G}}^<_{\bf p}+{\check \Sigma}^<_{\bf p}{\check {\rm
G}}^a_{\bf p}- {\check {\rm G}}^r_{\bf p} {\check \Sigma}^<_{\bf
p}-{\check {\rm G}}^<_{\bf p}{\check \Sigma}^a_{\bf p}).\label{CT}
\end{equation}
${\check {\rm G}}^{r,a,<}_{\bf p}$ and ${\check
\Sigma}^{r,a,<}_{\bf p}$ are, respectively, the nonequilibrium
Green's functions and self-energies. For brevity, hereafter, the
argument $({\bf p},\omega)$ of these functions will be denoted by
a subscript ${\bf p}$. In the kinetic equation (\ref{KE}) above, the
hole-impurity scattering is embedded in the self-energies,
${\check \Sigma}^{r,a,<}_{\bf p}$. In present paper, we
consider hole-impurity collisions only in the self-consistent Born
approximation. It is widely accepted that this is sufficiently
accurate to analyze transport properties in the diffusive regime.
Accordingly, the self-energies take the forms: ${\check
\Sigma}^{r,,a,<}_{\bf p}=n_i\sum_{\bf k} |V({\bf p}-{\bf k})|^2
{\check {\rm G}}_{\bf k}^{r,a,<}$, with impurity density $n_i$.

It is most convenient to study the hole distribution function in
the helicity basis, $\hat {\rho}({\bf p})= U^+({\bf p})\check
{\rho }({\bf p}) U({\bf p})$, because, there, the unperturbed
equilibrium distribution and the equilibrium lesser, retarded, and
advanced Green's functions are all diagonal. To derive the kinetic
equation for the helicity-basis distribution, ${\hat \rho}({\bf
p})$, we multiply Eq.\,(\ref{KE}) from left by $U_{\bf p}^+$ and
from right by $U_{\bf p}$. Due to the unitarity of $U_{\bf p}$,
the collision term in the helicity basis, $\hat I$, has a form
similar to Eq.\,(\ref{CT}), but with the helicity-basis Green's
functions and self-energies, $\hat {\rm G}^{r,a,<}_{{\bf
p}}=U^+({\bf p})\check {\rm G}^{r,a,<}_{{\bf p}}U({\bf p})$ and
$\hat {\Sigma}^{r,a,<}_{{\bf p}}=U^+({\bf p})\check
{\Sigma}^{r,a,<}_{{\bf p}}U({\bf p})$, respectively, replacing
those of the spin-basis, $\check {\rm G}^{r,a,<}_{{\bf p}}$ and
$\check {\Sigma}^{r,a,<}_{{\bf p}}$. The left hand side (LHS) of
Eq.\,(\ref{KE}) is simplified by using the following facts: $U_{\bf
p}^+\nabla_{\bf p} {\check \rho}({\bf p})U_{\bf p}=\nabla_{\bf
p}{\hat \rho}({\bf p}) -\nabla_{\bf p}U_{\bf p}^+U_{\bf p}{\hat
\rho}({\bf p})-{\hat \rho}({\bf p}) U_{\bf p}^+\nabla_{{\bf p}}
U_{\bf p}$ and $\nabla_{\bf p} U_{\bf p}^+U_{\bf p}=-U_{\bf
p}^+\nabla_{\bf p}U_{\bf p}$. Thus, the kinetic equation in
helicity basis may be written as
\begin{equation}
e{\bf E}\cdot \left \{\nabla_{\bf p} \hat \rho({\bf p})+[\hat
\rho({\bf p}), \nabla_{\bf p} U^+_{\bf p}U_{\bf p}] \right
\}+i[\hat h_0(p),\hat \rho({\bf p})]=-\hat I.
\end{equation}
In this equation, the helicity-basis self-energies, $\hat{\Sigma}^{r,a,<}_{\bf p}$, take the forms,
\begin{equation}
\hat{\Sigma}^{r,a,<}_{\bf p}=n_i\sum_{{\bf k}}\hat T({\bf
p},{\bf k})\hat{\rm G}^{r,a,<}_{\bf k}\hat T^+({\bf p},{\bf
k}).\label{SE}
\end{equation}

In this paper, we restrict our considerations to the linear
response regime. In connection with this, all the functions, such
as the nonequilibrium Green's functions, self-energies and
distribution, can be expressed as sums of two terms: $ A=
A_0+A_1$, with $ A$ as the Green's functions, self-energies or
distribution function. $A_0$ and $ A_1$, respectively, are the
unperturbed part and the linear electric field part of $A$. In
this way, the kinetic equation for the linear electric field part
of the distribution, $\hat {\rho}_1({\bf p})$, can be written as
\begin{equation}
e{\bf E}\cdot \nabla_{\bf p}\hat \rho_0({\bf p})-e{\bf E}\cdot
[\hat \rho_0({\bf p}), U_{\bf p}^+\nabla_{\bf p} U_{\bf p}]
+i[\hat h_0({\bf p}),\hat \rho_1({\bf p})]=-\hat
I^{(1)},\label{EQ11}
\end{equation}
with $\hat I^{(1)}$ as the linear electric field part of the
collision term $\hat I$:
\begin{eqnarray}
\hat I^{(1)}&=& \int \frac{d\omega}{2\pi}\left [{\hat
\Sigma}^r_{1\bf p}{\hat {\rm G}}^<_{0\bf p}+{\hat \Sigma}^<_{1\bf
p}{\hat {\rm G}}^a_{0\bf p}- {\hat {\rm G}}^r_{1\bf p} {\hat
\Sigma}^<_{0\bf p}-{\hat {\rm G}}^<_{1\bf
p}{\hat \Sigma}^a_{0\bf p}\right .\nonumber\\
&& \left .+{\hat \Sigma}^r_{0\bf p}{\hat {\rm G}}^<_{1\bf p}+{\hat
\Sigma}^<_{0\bf p}{\hat {\rm G}}^a_{1\bf p}- {\hat {\rm
G}}^r_{0\bf p} {\hat \Sigma}^<_{1\bf p}-{\hat {\rm G}}^<_{0\bf
p}{\hat \Sigma}^a_{1\bf p}\right ].
\end{eqnarray}

Further, we employ a two-band generalized Kadanoff-Baym ansatz
(GKBA)\cite{GKBA,GKBA1} to simplify Eq.\,(\ref{EQ11}). This
ansatz, which expresses the lesser Green's function through the
Wigner distribution function, has been proven sufficiently
accurate to analyze transport and optical properties in
semiconductors.\cite{Jauho} To first order in the dc field
strength, the GKBA reads,
\begin{equation}
    \hat {\rm G}^<_{1{\bf p}}=-\hat {\rm G}_{0{\bf p}}^r\hat \rho_1({\bf p})+\hat \rho_1({\bf p})\hat {\rm G}_{0{\bf p}}^a
    -\hat {\rm G}_{1{\bf p}}^r\hat \rho_0({\bf p})+\hat \rho_0({\bf p})\hat {\rm G}_{1{\bf p}}^a,\label{GKBA1}
\end{equation}
where the equilibrium distribution, and retarded and advanced
Green's functions are all diagonal matrices: $\hat \rho_0({\bf
p})={\rm diag}[n_{\rm F}(\varepsilon_H(p)),n_{\rm
F}(\varepsilon_L(p)),n_{\rm F}(\varepsilon_L(p)),n_{\rm
F}(\varepsilon_H(p))]$ and $\hat {\rm G}_0^{r,a}({\bf p})={\rm
diag}[(\omega-\varepsilon_H(p)\pm
i\delta)^{-1},(\omega-\varepsilon_L(p)\pm
i\delta)^{-1},(\omega-\varepsilon_L(p)\pm
i\delta)^{-1},(\omega-\varepsilon_H(p)\pm i\delta)^{-1}]$, with
the Fermi function $n_{\rm F}(\omega)$. We note that $\hat {\rm
G}_{1{\bf p}}^{r,a}$ in the collision term leads to a collisional
broadening of the nonequilibrium distribution. In the present
transport study, such collisional broadening plays a secondary
role and can be ignored. Based on this, the collision term, $\hat
I^{(1)}$, no longer involves the linear electric field part of the
retarded and advanced Green's functions.

It is obvious that the driving force in Eq.\,(\ref{EQ11})
comprises two components: the first of which, $e{\bf
E}\cdot\nabla_{\bf p}\hat\rho_0$, is diagonal, while another one,
$-e{\bf E}\cdot [\hat \rho_0({\bf p}), U_{\bf p}^+\nabla_{\bf p}
U_{\bf p}]$, has null diagonal elements. In connection with this,
we formally split the kinetic equation into two equations with
$\hat \rho_1({\bf p})=\hat \rho_1^{I}({\bf p})+\hat
\rho_1^{II}({\bf p})$ as
\begin{equation}
e{\bf E}\cdot \nabla_{\bf p} \hat \rho_0({\bf p})+i[\hat h_0({\bf
p}),\hat \rho_1^I({\bf p})]=-\hat I^{(1)},\label{EQ1}
\end{equation}
\begin{equation}
   -e{\bf E}\cdot [\hat \rho_0({\bf p}), U_{\bf p}^+\nabla_{\bf p} U_{\bf p}]
    +i[\hat h_0({\bf p}),\hat \rho_1^{II}({\bf p})]=0,\label{EQ2}
\end{equation}
wherein $\hat \rho_1^I({\bf p})$ and $\hat \rho_1^{II}({\bf p})$
can be approximately determined independently, as discussed below.
We note that the solution of Eq.\,(\ref{EQ2}), $\hat
\rho_1^{II}({\bf p})$, is off-diagonal and independent of impurity
scattering. The matrix elements of $\hat \rho_1^{I,II}({\bf p})$ will
be denoted by $(\hat \rho_1^{I,II})_{\mu\nu}({\bf p})$, and from Eqs.\,(\ref{JYX})
and (\ref{JYXE}), we correspondingly write spin-Hall conductivity
contributions based on $J_y^x=\left . J_y^x\right |^I+\left . J_y^x\right |^{II}$
as
\begin{eqnarray}
(\sigma^I)_{yz}^x=\left . J_y^x\right |^I/E=
\sum_{{\bf p}}{\rm Tr}[{\hat j}_y^x({\bf p}){\hat \rho}_1^I({\bf p})];
\nonumber\\
(\sigma^{II})_{yz}^x=\left . J_y^x\right |^{II}/E=
\sum_{{\bf p}}{\rm Tr}[{\hat j}_y^x({\bf p}){\hat \rho}_1^{II}({\bf p})].
\end{eqnarray}

It is evident that the diagonal driving term of Eq.\,(\ref{EQ1}),
$e{\bf E}\cdot\nabla_{\bf p}\hat\rho_0$, is free of impurity
scattering. Since $[\hat h_0,\hat \rho_1^I({\bf p})]$ is
off-diagonal, the diagonal parts of this equation lead to diagonal
$\hat \rho_1^I({\bf p})$ elements, $(\hat \rho_1^I)_{\mu\mu}({\bf
p})$ ($\mu=1...4$), of order of $(n_i)^{-1}$ in the impurity
density. Substituting these diagonal elements, $(\hat
\rho_1^I)_{\mu\mu}({\bf p})$, into the off-diagonal elements of
the scattering term, $\hat I^{(1)}$, and considering the fact that
the terms on LHS of the off-diagonal components of
Eq.\,(\ref{EQ1}) are proportional to the off-diagonal elements of
$\hat \rho_1^I({\bf p})$, we find that the leading order of the
off-diagonal elements of $\hat \rho_1^I({\bf p})$ in the
impurity-density expansion is of order $(n_i)^0$, {\it i.e.} independent of
$n_i$. This result implies that, in general, there always exists a
contribution to the spin-Hall current which is disorder-related
but independent of impurity density within the diffusive regime.
On the other hand, the off-diagonal impurity-density-independent
$\hat \rho_1^I({\bf p})$ elements, as well as all the nonvanishing
elements of $\hat \rho_1^{II}({\bf p})$, make contributions to the
scattering term, $\hat I^{(1)}$, which are linear in the impurity
density, while the $\hat I^{(1)}$ terms involving diagonal
elements, $(\hat \rho_1^I)_{\mu\mu}({\bf p})$, are independent of
$n_i$. Hence, the contributions to $\hat I^{(1)}$ from
off-diagonal elements of $\hat \rho_1({\bf p})$ can be ignored and
$\hat I^{(1)}$ effectively involves only the diagonal elements of
the distribution. Correspondingly, Eqs.\,(\ref{EQ1}) and
(\ref{EQ2}) are approximately independent of each other and can be
solved separately.

\subsection{Disorder-independent spin-Hall effect}

The disorder-independent spin-Hall current is associated with
$\hat \rho_1^{II}({\bf p})$, the solution of Eq.\,(\ref{EQ2}). The
nonvanishing elements of this function are given by
\begin{eqnarray}
(\hat\rho_1^{II})_{12}({\bf p})&=&-(\hat\rho_1^{II})_{21}({\bf p})
=(\hat\rho_1^{II})_{34}({\bf p})=-(\hat\rho_1^{II})_{43}({\bf p})\nonumber\\
&=&\frac{\sqrt{3}m}{4\gamma_2p^3}ieE\sin\theta_{\bf
p}[f_0^H(p)-f_0^L(p)],
\end{eqnarray}
with $f_0^H(p)=n_{\rm F}[\varepsilon_{H}(p)]$ and $f_0^L(p)=n_{\rm
F}[\varepsilon_{L}(p)]$, while its remaining elements, such as
$(\hat \rho_1^{II})_{13}({\bf p})$, $(\hat\rho_1^{II})_{24}({\bf
p})$, {\it etc.} vanish. Substituting $\hat\rho_1^{II}({\bf p})$
into Eq.\,(\ref{JYXE}), we find that the disorder-independent
contribution to intrinsic spin-Hall current, $\left .J_y^x\right
|^{II}$, can be written as
\begin{equation}
\left .J_y^x\right
|^{II}=\frac{eE}{6\pi^2}\int_0^\infty[f_0^H(p)-f_0^L(p)]dp.\label{JYX2}
\end{equation}
This result agrees with that obtained in
Ref.\,\onlinecite{Zhang2}.

Obviously, the nonvanishing of $\left .J_y^x\right |^{II}$ is
associated with the nonzero driving term on LHS of
Eq.\,(\ref{EQ2}), which is just the interband electric dipole
moment between the heavy- and light-hole bands. Thus, the
disorder-independent spin-Hall effect arises essentially from the
polarization process between two hole bands directly induced by
the dc electric field. Such a polarization can also be interpreted
as a two-band quantum interference process. It should be noted
that this polarization process affects only those off-diagonal
$\hat \rho^{II}_1({\bf p})$ elements which describe dc-field
induced transitions between hole states in the light- and
heavy-hole bands. Of course, such transition processes are not
restricted only to hole states in the vicinity of the Fermi
surface: they contribute from all the hole states below the Fermi
surface. As a result, the disorder-independent spin-Hall current
given by Eq.\,(\ref{JYX2}) is a function of the entire unperturbed
equilibrium distribution, $n_{\rm F}(\omega)$, not just of its
derivative, $\partial n_{\rm F}(\omega)/\partial \omega$, at the
Fermi surface.

\subsection{disorder-related spin-Hall effect}

To simplify Eq.\,(\ref{EQ1}), we first analyze symmetry relations
between the elements of the distribution function $\hat
\rho^I_{1}({\bf p})$ in the self-consistent Born approximation.
Since the distribution function is a Hermitian matrix, only the
independent elements $(\hat \rho^I_{1})_{\mu\nu}({\bf p})$ with
$\mu,\nu=1...4$ and $\mu\le \nu$ need to be considered. We know
that $(\hat \rho^I_1)_{11}({\bf p})$ and $(\hat
\rho^I_1)_{44}({\bf p})$ describe the distributions of the heavy
holes having spins $S_z=3/2$ and $S_z=-3/2$, respectively. In
equilibrium, heavy hole populations in degenerate states with
$S_z=3/2$ and $S_z=-3/2$ distribute equally. Out of equilibrium,
the dc electric field action on these hole populations is also the same.
Hence, the nonequilibrium distribution of the heavy holes with
$S_z=3/2$ is the same as that of the heavy holes with
$S_z=-3/2$, {\it i.e.} $(\hat \rho_{1}^{I})_{11}({\bf p})=(\hat
\rho_{1}^{I})_{44}({\bf p})$. An analogous relation for light
holes is also expected to be valid: $(\hat \rho_{1}^{I})_{22}({\bf
p})=(\hat \rho_{1}^{I})_{33}({\bf p})$. Indeed, substituting these
symmetrically related diagonal elements of the distribution $\hat
\rho^I_1({\bf p})$ into the scattering term, we find $\hat
I^{(1)}_{11}=\hat I^{(1)}_{44}$, $\hat I^{(1)}_{22}=\hat
I^{(1)}_{33}$, and $\hat I^{(1)}_{23}=\hat I^{(1)}_{32}=\hat
I^{(1)}_{14}=\hat I^{(1)}_{41}=0$, which are consistent with the
elements on the LHS of Eq.\,(\ref{EQ1}). As another consequence of
these relations ($(\hat \rho_{1}^{I})_{11}({\bf p})=(\hat
\rho_{1}^{I})_{44}({\bf p})$ and $(\hat \rho_{1}^{I})_{22}({\bf
p})=(\hat \rho_{1}^{I})_{33}({\bf p})$), we also obtain symmetry
relations between the remaining off-diagonal elements of $\hat
I^{(1)}$: $\hat I^{(1)}_{12}=-\hat I^{(1)}_{34}$ and $\hat
I^{(1)}_{13}=\hat I^{(1)}_{24}$, which result in symmetry
relations for the $\hat \rho_{1}^{I}({\bf p})$ elements as: $(\hat
\rho_{1}^{I})_{12}({\bf p})=(\hat \rho_{1}^{I})_{34}({\bf p})$ and
$(\hat \rho_{1}^{I})_{13}({\bf p})=-(\hat \rho_{1}^{I})_{24}({\bf
p})$. Hence, to determine the disorder-related spin-Hall effect,
one only needs to evaluate the diagonal elements, $(\hat
\rho_{1}^{I})_{11}({\bf p})$ and $(\hat \rho_{1}^{I})_{22}({\bf
p})$, and the off-diagonal elements, $(\hat
\rho_{1}^{I})_{12}({\bf p})$ and $(\hat \rho_{1}^{I})_{13}({\bf
p})$.

From Eq.\,(\ref{EQ1}), it follows that the diagonal $\hat
\rho_1^I({\bf p})$ elements are determined by the integral equation
\begin{eqnarray}
-e{\bf E}\cdot {\bf \nabla}_{\bf p}n_{\rm F}[\varepsilon_\mu (p)]&=&\pi\sum_{\bf k}|V({\bf p}-{\bf k})|^2
\{a_1({\bf p},{\bf k})[(\hat \rho_1^I)_{\mu\mu}({\bf p})\nonumber\\
&&-(\hat \rho_1^I)_{\mu\mu}({\bf k})]\Delta_{\mu\mu}+a_2({\bf p},{\bf k})[(\hat\rho_1^I)_{\mu\mu}({\bf p})
\nonumber\\
&&-(\hat\rho_1^I)_{\bar \mu\bar \mu}({\bf
k})\Delta_{\mu\bar\mu}].\label{KEE}
\end{eqnarray}
Here, $\mu=1,2$, respectively, correspond to the heavy- and
light-hole bands: $\varepsilon_1(p)\equiv \varepsilon_H(p)$,
$\varepsilon_2(p)\equiv \varepsilon_L(p)$, $\bar \mu=3-\mu$,
$\Delta_{\mu\nu}=\delta [\varepsilon_\mu (p)-\varepsilon_\nu
(k)]$. The factors $a_1({\bf p},{\bf k})$ and $a_2({\bf p},{\bf
k})$ are associated only with the momentum angles:
\begin{eqnarray}
a_1({\bf p},{\bf k})&=&\frac 14 \{2+6\cos^2 \phi_{pk}[\sin^2 \theta_{\bf p}-\cos^2 \theta_{\bf k}]
\nonumber\\
&&+6\cos^2\theta_{\bf p}\cos^2\theta_{\bf k}[1+\cos^2 \phi_{pk}]\nonumber\\
&&+3\cos \phi_{pk}\cos(2\theta_{\bf p})\cos(2\theta_{\bf k})\},
\end{eqnarray}
\begin{equation}
a_2({\bf p},{\bf k})=2-a_1({\bf p},{\bf k}),
\end{equation}
where $\phi_{pk}\equiv \phi_{\bf p}-\phi_{\bf k}$. From
Eq.\,(\ref{KEE}), we see that we may remove the dependence of
$(\rho_1^I)_{\mu\mu}({\bf p})$ on momentum angle $\phi_{\bf p}$ by
redefining the angular integration variable as $\phi_{{\bf
k}}\rightarrow \phi_{pk}=\phi_{\bf p}-\phi_{\bf k}$, taken jointly
with the facts that the left hand side does not depend on
$\phi_{\bf p}$ and the potential $V({\bf p}-{\bf k})$, as well as
the factors $a_1({\bf p},{\bf k})$ and $a_2({\bf p},{\bf k})$,
depends on $\phi_{\bf p}$ and $\phi_{\bf k}$ only through the
combination $\phi_{pk}$.

Analyzing the components of the scattering term in the kinetic
equation for the off-diagonal elements, $(\hat \rho_1^I)_{12}({\bf
p})$ and $(\hat \rho_1^I)_{13}({\bf p})$, we find that these
elements of the distribution $\hat \rho_1^I({\bf p})$ are
similarly effectively independent of $\phi_{\bf p}$. In connection
with this, contributions to the disorder-related spin-Hall
current, $\left .J_y^x\right |^I$, from $(\hat \rho_1^I)_{13}({\bf
p})$ and ${\rm Re} [(\hat \rho_1^I)_{12}({\bf p})]$ vanish under
the $\phi_{\bf p}$-integration in Eq.\,(\ref{JYXE}), and only the
imaginary part of $(\hat \rho_1^I)_{12}({\bf p})$ makes a
nonvanishing contribution to $\left .J_y^x\right |^I$. Hence,
\begin{equation}
\left .J_y^x\right|^I=\frac{8\sqrt{3}\gamma_2}{m}\sum_{\bf p}p\left \{\cos^2\phi_{\bf p}\sin \theta_{\bf p} {\rm Im}[(\hat \rho_{1}^I)_{12}({\bf p})]
\right \},\label{JYX1}
\end{equation}
with
\begin{eqnarray}
{\rm Im}\left [(\hat \rho_1^I)_{12}({\bf p})\right ]&=&\frac{\sqrt{3}\pi m}{4\gamma_2 p^2}\sum_{{\bf k},\mu=1,2}
|V({\bf p}-{\bf k})|^2a_3({\bf p},{\bf k})\nonumber\\
&&\times(-1)^\mu\{\Delta_{\mu\mu}[(\hat \rho_1^I)_{\mu\mu}({\bf p})-(\hat \rho_1^I)_{\mu\mu}({\bf k})]
\nonumber\\
&&-\Delta_{\mu\bar\mu}[(\hat \rho_1^I)_{\mu\mu}({\bf p})-(\hat \rho_1^I)_{\bar\mu\bar\mu}({\bf k})]\},\label{KEE1}
\end{eqnarray}
and
\begin{eqnarray}
a_3({\bf p},{\bf k})&=&-\frac 12 \{\sin(2\theta_{\bf p})[\cos^2\theta_{\bf k}
-\sin^2\theta_{\bf k}\cos^2\phi_{pk}]
\nonumber\\
&&+\sin(2\theta_{\bf k})\cos\phi_{pk}[1-2\cos^2\theta_{\bf p}]\}.
\end{eqnarray}

From Eqs.\,(\ref{KEE}) and (\ref{KEE1}), we see that $\left
.J_y^x\right |^I$ is independent of impurity density. In contrast
to the disorder-independent case, the disorder-related spin-Hall
current involves only the derivative of the equilibrium
distribution function, {\it i.e.} $\partial n_{\rm
F}(\omega)/\partial \omega$. This implies that $\left .J_y^x\right
|^I$ is constituted of contributions arising only from hole states
in the vicinity of the Fermi surface, or in other words, from hole
states involved in longitudinal transport. Physically, the holes
participating in transport experience impurity scattering,
producing diagonal $\hat \rho^I_1({\bf p})$ elements of order of
$n_i^{-1}$. Moreover, the scattering of these perturbed holes by
impurities also gives rise to an interband polarization, which no
longer depends on impurity density within the diffusive regime. It
is obvious that in such a polarization process the disorder plays
only an intermediate role. It should be noted that $\left
.J_y^x\right |^I$ generally depends on the form of the
hole-impurity scattering potential, notwithstanding its
independence of impurity density in the diffusive regime.

The fact that the total spin-Hall current, $J_y^x= \left
.J_y^x\right |^I+\left .J_y^x\right |^{II}$, consists of two
parts associated with hole states below and
near the Fermi surface, respectively, is similar to the well-known result of
St\v reda\cite{Streda} in the context of the 2D charge Hall
effect. In 2D electron systems in a normal magnetic field, the
off-diagonal conductivity usually arises from two terms, one of
which is due to electron states near the Fermi energy and the
other is related to the contribution of all occupied electron
states below the Fermi energy. A similar picture has also recently
emerged in studies of the anomalous Hall effect.\cite{AHE1,AHE2}

\section{Results and discussions}

To compare our results with the short-range result presented in
Ref.\,\onlinecite{Murakami}, we first consider a short-range
hole-impurity scattering potential described by: $V({\bf
p}-{\bf k})\equiv u$, with $u$ as a constant. Substituting
Eq.\,(\ref{KEE1}) into Eq.\,(\ref{JYXE}) and performing
integrations with respect to the angles of $\bf p$ or $\bf k$,
respectively, for terms involving $(\hat \rho_1^I)_{\mu\mu} ({\bf
k})$ or $(\hat \rho_1^I)_{\mu\mu} ({\bf p})$, we find that the
contribution of short-range disorder to the spin-Hall current
vanishes, {\it i.e.} $\left. J_y^x\right |^I=0$. This implies that
for short-range hole-impurity collisions, the total spin-Hall
current is just the disorder-independent one, $J_y^x=\left. J_y^x
\right |^{II}$. This result agrees with that obtained in
Ref.\,\onlinecite{Murakami}.

Furthermore, we perform a numerical calculation to investigate the
effect of long-range hole-impurity collisions on the spin-Hall
current in a GaAs bulk semiconductor. The long-range scattering is
described by a screened Coulombic impurity potential
$V(p)$: $V(p)=e^2/(\varepsilon_0\varepsilon) [p^2+1/d^2_D]^{-1}$
with $\varepsilon$ as a static dielectric constant.\cite{Grill}
$d_D$ is a Thomas-Fermi-Debye type screening length:
$d_D^2=\pi^2\varepsilon_0\varepsilon/(e^2\sqrt{2m^3E_{F}})2^{-1/3}
[(\gamma_1+2\gamma_2)^{-3/2}+(\gamma_1-2\gamma_2)^{-3/2}]^{-2/3}$,
with $E_F=(3\pi^2 N_p/2)^{2/3}/(2m)$. The material parameters
$\gamma_1$ and $\gamma_2$ are chosen to be $6.85$ and $2.5$,
respectively.\cite{GAAS} In our calculation, the momentum
integration is computed by the Gauss-Legendre scheme.

In the present paper, we address the spin-Hall effect at zero
temperature, $T=0$. In this case, the disorder-independent
spin-Hall current can be obtained analytically from
Eq.\,(\ref{JYX2}): $\left .J_y^x\right
|^{II}=eE[k_F^H-k_F^L]/(6\pi^2)$, with $k_F^H$ and $k_F^L$ as the
Fermi momenta for heavy- and light-hole bands, respectively. In
order to investigate the disorder-related spin-Hall effect, we
need to compute the distribution function $\hat \rho_1^I({\bf p})$
at the Fermi surface. In this calculation, we employ a "singular
value decomposition" method\cite{NR} to solve the integral
equation, Eq.\,(\ref{KEE}), for the diagonal $\hat \rho_1^I({\bf
p})$ elements. The obtained diagonal elements are then employed to
determine ${\rm Im}[(\rho_1^I)_{12}({\bf p})]$ using
Eq.\,(\ref{KEE1}). Following that, we obtain the disorder-related
spin-Hall current from Eq.\,(\ref{JYX1}), performing the momentum
integration.

\begin{figure}
\includegraphics [width=0.45\textwidth,clip] {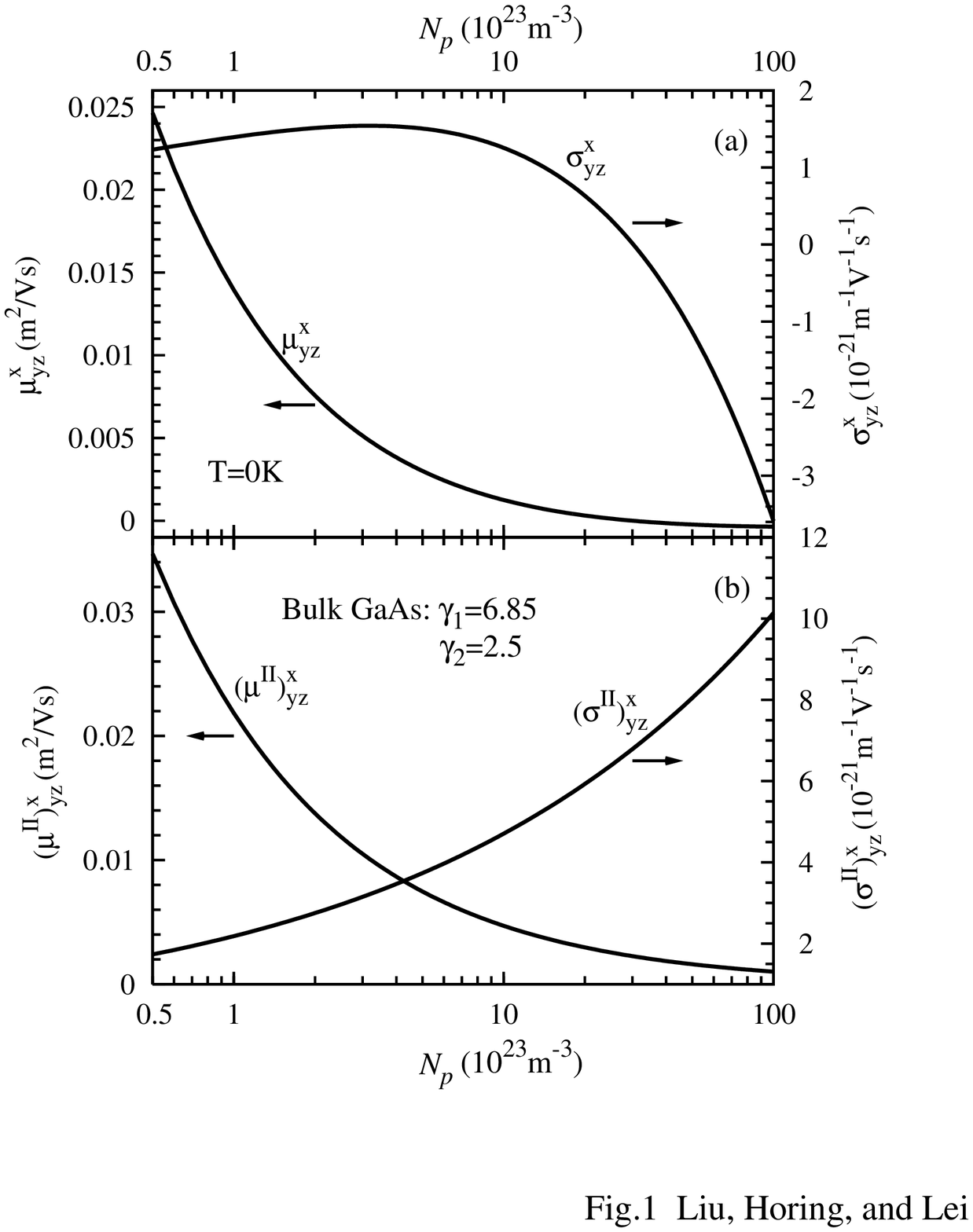}
\caption{Hole-density dependencies of (a) total $\sigma_{yz}^x$
and $\mu_{yz}^x$, and (b) disorder-independent
$(\sigma^{II})_{yz}^x$ and $(\mu^{II})_{yz}^x$, in a bulk GaAs
semiconductor. The material parameters for GaAs are:
$\gamma_1=6.85$ and $\gamma_2=2.5$. The lattice temperature is
$T=0$\,K.} \label{fig1}
\end{figure}

In Fig.\,1, the calculated total and disorder-independent
spin-Hall conductivities, $\sigma_{yz}^x=J_{y}^x/E$ and
$(\sigma^{II})_{yz}^x=\left .J_{y}^x\right |^{II}/E$, and the total
and disorder-independent spin-Hall mobilities,
$\mu_{yz}^x=\sigma_{yz}^x/N_{p}$ and $(\mu^{II})_{yz}^x
=(\sigma^{II})_{yz}^x/N_{p}$, are shown as functions of the hole
density. The spin-Hall mobility, analogous to the mobility of
charge transport, characterizes the average mobile ability of a
single spin driven by the external field. This quantity has the
same units in 2D and 3D systems.

From Fig.\,1, we see that, with increasing hole density, the total
spin-Hall conductivity first increases and then decreases and even
becomes negative as the hole density becomes larger than
$N_{pc}=3\times 10^{24}$\,m$^{-3}$. This behavior of the
hole-density dependence of total spin-Hall conductivity is the
result of a competition between the disorder-independent and
disorder-related processes. The contributions to spin-Hall
conductivity from these two processes always have opposite signs
and their absolute values increase with increasing hole density.
Considering total spin-Hall conductivity, the disorder-related
part, $(\sigma^{I})_{yz}^x$, is dominant for high hole density,
while $(\sigma^{II})_{yz}^x$ is important in the low hole-density
regime. Notwithstanding this hole-density dependence of
$\sigma_{yz}^x$, the total spin-Hall mobility, $\mu_{yz}^x$, as
well as the disorder-independent one, monotonically decreases with
increasing hole density.

It should be noted that the total spin-Hall mobility in bulk
systems has the same order of magnitude as that in 2D hole
systems. We know that the spin-Hall conductivity in 2D hole
systems is of order of $e/\pi$.\cite{Liu3} For a typical 2D hole
density, $n_p^{(2D)}=1\times 10^{12}$\,cm$^{-2}$, the
corresponding spin-Hall mobility is about $0.05$\,m$^2$/Vs.

In the present paper, we have ignored the effect of collisional
broadening on spin-Hall current. Since $\left . J_y^x\right |^I$
is associated only with the hole states in the vicinity of the
Fermi surface, the neglect of broadening in the disorder-related
spin-Hall current is valid for $\varepsilon_{F} \tau> 1$
($\varepsilon_F$ is the Fermi energy and $\tau$ is the larger of
the relaxation times for holes in the different bands at the Fermi
surface, $\tau_L(\varepsilon_F)$ and $\tau_H(\varepsilon_F)$:
$\tau=\max [\tau_L(\varepsilon_F),\tau_H(\varepsilon_F)]$). This
condition coincides with the usual restriction on transport in the
diffusive regime and is satisfied for $p$-type bulk GaAs with
mobility approximately larger than $1$\,m$^2$/Vs (for $N_{\bf
p}>5\times 10^{22}$\,m$^{-3}$). On the other hand, the
disorder-independent spin-Hall conductivity involves contributions
from all hole states in the Fermi sea and hence it may be strongly
affected by collisional broadening. To estimate the broadening
effect on the disorder-independent SHE, we add an imaginary part
to $\hat h_0({\bf p})$ and use $\hat h_0({\bf p})+i\hat
\gamma({\bf p})$ instead of $\hat h_0({\bf p})$ in
Eq.\,(\ref{EQ2}) ( $\hat \gamma({\bf p})$ is a diagonal matrix
describing the broadening: $(\hat \gamma)_{11}({\bf p})=(\hat
\gamma)_{44}({\bf p})=1/2\tau_H(\varepsilon_H(p))$ and $(\hat
\gamma)_{22}({\bf p})=(\hat \gamma)_{33}({\bf
p})=1/2\tau_L(\varepsilon_L(p))$). In this way, $\left.
J_y^x\right |^{II}$ takes a form similar to Eq.\,(\ref{JYX2}) but
with an additional factor, $(2\gamma_2 p^2)^2/\{[2\gamma_2
p^2]^2+[1/2\tau_H(\varepsilon_H(p))-1/2\tau_L(\varepsilon_L(p))]^2\}$,
in the momentum integrand. Performing a numerical calculation, we
find that, in the studied regime of hole density, the effect of
collisional broadening on the disorder-independent spin-Hall
current is less than 1\% for $p$-type bulk GaAs semiconductors
with mobility approximately larger than 5\,m$^2$/Vs. Thus, in such
systems, the effect of collisional broadening on the total
spin-Hall conductivity can be ignored. It should be noted that in
our calculations, we computed $\tau_{L,H}(\varepsilon)$ by
considering short-range hole-impurity scattering:
$1/\tau_{L,H}(\varepsilon)=2\pi n_iu^2\nu_{L,H}(\varepsilon)$ with
the densities of hole states in the light- and heavy-hole bands
taken as $\nu_{L,H}(\varepsilon)=2\sum_{\bf
p}\delta(\varepsilon-\varepsilon_{L,H}(p))$. The quantity $n_iu^2$
is determined from the mobility of the system:
$\mu=e[N_{p}^L\tau_L(\varepsilon_F)/m_L+N_p^H\tau_H(\varepsilon_F)/m_H]/N_p$,
where $m_L=m/(\gamma_1+2\gamma_2)$ and
$m_H=m/(\gamma_1-2\gamma_2)$ are the effective masses of holes and
$N_p^L/N_p^H=[(\gamma_1-2\gamma_2)/(\gamma_1+2\gamma_2)]^{3/2}$
with $N_p^L$ and $N_p^H$ being the hole densities in the light-
and heavy-hole bands, respectively.

On the other hand, in our considerations, the impurities are taken
to be so dense that we can use a statistical average over the
impurity configuration. This requires that $L_D<L$ ($L$ is the
characteristic size of the sample and $L_D$ is the larger of the
diffusion lengths of holes in the light- and heavy-hole bands).
Failing this, the behavior of the holes would become ballistic,
with transport properties depending on the specific impurity
configuration.

\section{Conclusions}
We have employed a nonequilibrium Green's function kinetic
equation approach to investigate disorder effects on the spin-Hall
current in the diffusive regime in $p$-type bulk Luttinger
semiconductors. Long-range hole-impurity scattering has been
considered within the framework of the self-consistent Born
approximation. We have found that, in contrast to the null effect
of short-range disorder on the spin-Hall current, long-range
scattering produces a nonvanishing contribution to the spin-Hall
current, independent of impurity density in the diffusive regime.
This contribution has its sign opposite to that of the
disorder-independent one, leading to a sign change of the total
spin-Hall current as the hole density varies. We also made clear
that the disorder-independent spin-Hall effect arises from a
dc-field-induced polarization associated with all hole states in the
Fermi sea, while the disorder-related one is produced by a
disorder-mediated polarization and relates to only those hole
states in the vicinity of the Fermi surface. The numerical
calculation indicates that with increasing hole density, the
total spin-Hall mobility monotonically decreases, whereas the
spin-Hall conductivity first increases and then falls.

In addition to $J_y^x$, we also examined other components of the
spin current. We found that the previously discovered "basic
spintronics relation",\cite{Zhang} which relates the $i$th
component of the spin current along the direction $j$, $J_j^i$,
and the applied electric field, $E_k$, by
$J_j^i=\sigma_s\epsilon_{ijk}E_k$ with $\epsilon_{ijk}$ as a
totally antisymmetric tensor, still holds in the presence of
spin-conserving hole-impurity scattering.

\begin{acknowledgments}
This work was supported by the Department of Defense through the
DURINT program administered by the US Army Research Office, DAAD
Grant No. 19-01-1-0592, and by projects of the National Science
Foundation of China and the Shanghai Municipal Commission of
Science and Technology.
\end{acknowledgments}

\end{document}